\documentstyle[prb,aps,multicol,epsfig,amsmath,amsfonts,psfrag,rotating,pst-feyn,pst-key]{revtex}
 \newcommand{\nts}[1]{\tmspace{-}{#1\thinmuskip}{#1\txtmu}}
 \newcommand\eqlabel[1]{\label{#1}}

\allowdisplaybreaks
\begin{document}
\draft

\title{The asymptotic behaviour
  of the exact and approximative \boldmath $ \nu=1/2 \; $ \unboldmath
  Chern-Simons Green's functions}
  
\author{J. Dietel}
\address{
Institut f\"ur Theoretische Physik, Universit\"at Leipzig,\\
Augustusplatz 10, D 04109 Leipzig, Germany }
\date{\today}
\maketitle
\begin{abstract}
We consider the asymptotic behaviour of the Chern-Simons Green's function
of the
$\nu=1/\tilde{\phi} $ system for an infinite area in position-time
representation. We calculate explicitly the
asymptotic form of the Green's function of the interaction free Chern-Simons
system for small times. The calculated Green's function
vanishes exponentially with the
logarithm of the area. Furthermore, we discuss the form of the divergence
for all $ \tau $ and also for the Coulomb interacting Chern-Simons system.
We compare the asymptotics of the exact Chern-Simons
Green's function with the asymptotics of the Green's function in the
Hartree-Fock
as well as the random-phase approximation (RPA).
The asymptotics of the Hartree-Fock Green's function corresponds well
with the exact Green's function. In the case of the
RPA Green's function we do not get the correct asymptotics.
At last, we calculate the
self consistent Hartree-Fock Green's function. 
\end{abstract}

\pacs{71.10.Pm, 73.43.-f, 71.27.+a}

\begin{multicols}{2}
\narrowtext

\section{Introduction}
The combination of an electronic interaction and a strong magnetic 
field in a two-dimensional electron system yields a rich variety of 
phases. These are best classified by the filling factor $ \nu $, 
which is the electron density divided by the density of a completely 
filled Landau level. In this paper, we consider systems with filling fraction
$ \nu=1/\tilde{\phi} $ where $\tilde{\phi}$ is an even positive number. 
These system are most suitably described
by the Chern-Simons theory. 
Since the discovery of the fractional quantum Hall effect 
by Tsui, St\"ormer and Gossard (1982) \cite{ts1} there were many attempts to  
explain this experimental observation.
The contemporary theoretical picture of this 
effect is mainly based on a work of Jain (1989) \cite{ja1}.    
In his theory he mapped the wave functions of the integer quantum Hall effect
to wave functions of the fractional quantum Hall effect. In the case of
filling fraction $ \nu=1/2 $ every electron gets two magnetic flux quantums
through  this mapping. With the help of
this transformation new quasi-particles
(composite fermions) are obtained which do not see any
magnetic field in first approximation (mean field).
A field theoretical language for this scenario was first established by
Halperin, Lee, Read (HLR) (1992) \cite{hlr}, as well as Kalmeyer and Zhang
(1992) \cite{ka1} for the $ \nu=1/2 $ system. The interpretation of
many experiments supports
this composite fermion picture. We mention transport experiments with
quantum (anti-) dots \cite{kan1}, and focusing experiments \cite{sm1} here.
An overview of further experiments can be found in \cite{wil2}.

HLR studied many physical quantities within the
random-phase approximation (RPA). Most prominent among these is the effective
mass of the composite fermions which they found to diverge at the Fermi
surface \cite{hlr,st1}. This is based on the interaction of the composite
fermions via transversal gauge interactions. Later on, Shankar and
Murthy \cite{sh1} proposed a
new theory of the $ \nu=1/2 $ system. Based upon a transformation
of the Chern-Simons Hamiltonian one achieves a separation
of the magneto-plasmon oscillators from the total interaction of the
system. After restricting the number of the magneto-plasmon oscillators
to the number of electrons Shankar and Murthy  got a finite
quasi-particle mass which scales with the inverse of the strength of the
Coulomb repulsion. In their derivation they calculated a smaller number of
self energy Feynman diagrams than in the RPA. Recently
Stern et al. \cite{st2} calculated the self energy 
of the theory of Shankar and Murthy in RPA finding
the same divergence of the
effective mass as HLR. Besides the theories of HLR and Shankar and Murthy
there are other alternative formulations of the Chern-Simons theory
which looks similar
to the Chern-Simons theory of Shankar and Murthy \cite{pas1}.

In this paper, we consider the asymptotic behaviour  of the
$ \nu=1/\tilde{\phi} $ Chern-Simons Green's function for an
infinite area $ A $ non-perturbationally.
It is well known that the Hartree-Fock approximation 
\cite{sit2} as well as the RPA \cite{hlr}  of the Chern-Simons
Green's function results in a $ \log(A) $ singularity  
in the momentum-frequency representation ($ \vec{q}$, $ \omega$)
of the self energy. This singularity is caused by the interaction of the
composite fermions
through a longitudinal gauge field (in contrast to the effective mass).
In almost all calculations of the effective mass of the
composite fermions this singularity is not taken in to account
although one can easily show that this  $ \log(A) $ singularity
enforces the effective mass to be finite (the bare mass).
Furthermore, this singularity of the self energy is not only given on
the fermi-surface
but for all momenta (apart from the fermi-surface).
The neglection of the $ \log(A) $ singularity
in the calculation of the effective mass is justified by the physical
argument that it must scale with  the inverse of the
Coulomb interaction \cite{hlr}.
By neglecting this singularity one gets the correct scaling.
This was the reason that till now almost all authors disregard
the $ \log(A) $ singularity and try to get rather a
better physical insight into the effective
mass singularity caused by the transverse gauge interaction (\cite{sim1} and
references therein). From the point of view that the Chern-Simons theory
is a many-body theory for which physical quantities are calculated by
known perturbative methods (although no small parameter is present)
this procedure is not satisfactory. On the way to integrate this
singularity in a Chern-Simons perturbation theory or to formulate a
Chern-Simons theory without this Green's function divergence, we will
investigate the $ \log(A) $ singularity in this paper.
The aim of our investigation is to get rather exact statements about this
singularity and the relation of perturbative
calculated Green's functions to the exact Green's function.

HLR \cite{hlr,ha2} gave in their  paper a semi-classical reason for the
$ \log(A) $ singularity
by showing that the Chern-Simons transformation effectively gives 
a velocity boost to every electron. This velocity boost results in a
one particle energy which diverges proportional to  $ \log(A) $.
Unfortunately, the derivation of HLR is based on semi-classical
approximations. To our knowledge there are no publications which show 
without approximations that the $ \log(A) $ singularity is really existent
in the Green's function or even how it looks like. 
In this paper, we will calculate the asymptotics of the Chern-Simons
Green's function in the position-time ($ \vec{r}$, $ \tau$) representation.
The asymptotics will be computed concretely for the
interaction free
$ \nu=1 / \tilde{\phi} $ Chern-Simons system for all $ \vec{r} $ and small
$ \tau $. We will show exactly that the Green's function vanishes for
$ A \to \infty $. 
Furthermore, we will discuss the form of the divergence
for all $ \tau $ and also for the Coulomb interacting Chern-Simons system.
It is clear that perturbational calculations of physical quantities
such as the energy should start with Green's and
vertex functions which are in rather good agreement with the exact
functions. To obtain a good approximation of the Chern-Simons
Green's function, we will calculate it in two
different approximations.
First, we will calculate the Hartree-Fock Green's
function. We will show that the asymptotic behaviour of
this Green's function is in 
good agreement with the asymptotic behaviour of the exact Green's function.
Then we will calculate the Chern-Simons Green's function in RPA.
We will show that this Green's function is finite for
$ A \to \infty $ (modulo logarithmic singularities), which is not in
agreement with the exact Green's
function. On the way to formulate a perturbation theory around the
Hartree-Fock mean field, we will examine at last the self consistent
Hartree-Fock Green's function.

In the following derivation of the Chern-Simons Green's function, we will
keep the formulas as general as possible. This will be done to simplify
the extension of  the
concrete calculation of the asymptotics of the Green's function
$ G(\vec{r},\tau) $ also for large $ \tau $ in a later publication.
In this publication, we restrict our calculation of the Green's function to
the range of small $ \tau $ because this $ \tau $ range is most
relevant for a calculation of  physical quantities  
(the Green's function
decreases by physical arguments for larger $ \tau $).
Furthermore, we restrict our calculations to the case temperature $ T=0 $.

 The paper is organized as follows:\\
In section II, we will calculate the asymptotic behaviour of the
Chern-Simons Green's function $ G(\vec{r},\tau) $.
In section III, we will calculate the self consistent Chern-Simons Green's
function in Hartree-Fock approximation as well as in RPA 
and compare these approximations with
the exact Chern-Simons Green's function calculated in section II.

\section{The asymptotic behaviour of the Chern-Simons Green's function} 

In this paper  
we consider interacting spin polarized electrons moving in two dimensions in a
strong magnetic field $ B $ directed in the negative  $ z $-direction
of the system.
The electronic density of the system is chosen such that the lowest Landau
level of a non-interacting system is filled to a fraction
$ \nu=1/ \tilde{\phi}$ where $ \tilde{\phi} $ is an even number. We are mainly
interested in $ \tilde{\phi}=2 $. The composite
fermion annihilation operator $ \Psi(\vec{r}) $ is defined from the 
electronic annihilation operator $ \Psi_e(\vec{r}) $
through the  Chern-Simons
transformation \cite{zh2}
($ \alpha(\vec{r}) $ is the angle
between the x-axes and the position $ \vec{r} $) 
\begin{equation}\eqlabel{1070}
\Psi(\vec{r})=e^{-i\phi \int d^2r' \alpha(\vec{r}-\vec{r}\,') 
\Psi^+(\vec{r}\,')\Psi(\vec{r}\,')}\Psi_e(\vec{r})
\end{equation}
The Hamiltonian of the composite fermions is then given by :
\begin{eqnarray}
& & H_{CS}=\int d^2r
\bigg[\frac{1}{2m}\big|\big(-i\vec{\nabla}+\vec{A}  
+ \vec{a}_{CS}\big)\Psi(\vec{r})\big|^2 \eqlabel{1050}   \\
& &   +\frac{1}{2} \nts{1} \int \nts{2} d^2r' \Big\{
 (|\Psi(\vec{r})|^2-\rho_B)
 V^{ee}(|\vec{r}-\vec{r}\,'|)(|\Psi(\vec{r}\,')|^2-\rho_B)\Big\}\bigg].
 \nonumber 
\end{eqnarray}
The Chern-Simons vector potential  $ \vec{a}_{CS} $ is defined
by $ \vec{a}_{CS}(\vec{r})=
\tilde{\phi} \int d^2r' \;\vec{f}(\vec{r}-\vec{r}\,')
\Psi^+(\vec{r}\,')\Psi(\vec{r}\,') $. 
$ m $ is the mass of an electron. 
$ V^{ee}(r)=e^2/r $ is the Coulomb interaction
where $ e^2=q_e^2 /\epsilon  $. $ q_e $ is the charge of the electrons and
$ \epsilon $ is the dielectric constant of the background field $ \rho_B $.
$ \vec{A}(\vec{r}) $ is the vector potential \
$ \vec{A}=1/2 \,\vec{B} \times \vec{r} $ and $\vec{B} $ is a 
homogeneous magnetic field in the negative z-direction
$ \vec{B}=-B \vec{e}_z $,
where $ \vec{e}_z $ is the
unit vector in $z $-direction. 
We suppose throughout this paper that $ B $ is a positive number. 
The function $ \vec{f}(\vec{r}) $ is given by 
$ \vec{f}(\vec{r})=\vec{e}_z \times \vec{r}/r^2 $.
We used 
the convention $ \hbar=1 $ and $ c=1$ in the above formula
(\ref{1050}). Furthermore, we set $ q_e=1 $ for the coupling of the magnetic
potential to the electrons.
The Green's function of the Chern-Simons theory is defined as 
\begin{equation}\eqlabel{1515}
G(\vec{r},t;\vec{r}\,',t')=
\langle T[\Psi(\vec{r},t),\Psi^+(\vec{r}\,',t')]\rangle_{CS} \;.
\end{equation}
$ \langle \cdot \rangle_{CS} $ is the average over the Gibb's operator of $
H_{CS} $. $ T $ is the time ordering operator.
Applying  the inverse of the Chern-Simons transformation (\ref{1070})
to the operator $ H_{CS} $, we get the electronic Hamiltonian $ H_{e} $.
$ H_{e} $ is given by $ H_{CS} $
(\ref{1050}) by the substitutions   $ a_{CS}=0 $, $ \Psi \to \Psi_e $ and
$ \Psi^+ \to \Psi_e^+ $.

As mentioned in the introduction, we will at first calculate the $ N $
particle Green's
function for $ V_{ee}=0 $.
Under this restriction the ground state of 
$ H_{e} $ is degenerate. The ground state wave functions are given by 
\begin{equation}\eqlabel{1517}
u_{0,\vec{p}}=S[u_{p_1},u_{p_2}, 
\cdots, u_{p_N}] \;.
\end{equation}
$ S$ is the Slater determinant of the one particle wave functions 
$ u_{p_1},.. ,u_{p_N}$.
$ u_{p} $ are the one particle wave functions in the lowest Landau level 
in the symmetric gauge 
\begin {equation}\eqlabel{1518}
u_{p}(r,\phi)=\sqrt{\frac{1}{2^{p+1}\;p!\; \pi}}\;e^{ip\phi}\; r^p 
\;e^{-\frac{1}{4}r^2} \;.
\end{equation}
In equation (\ref{1517}), we have $ p \in 0 \ldots \tilde{\phi} N $
(for finite $ A $).
Further, we use the simplifications $ B=1 $, $m=1 $. This will be
done throughout this section in the auxiliary equations.
For equations
which have the character of a result, we will insert the $ B $ and $ m $
dependencies explicitly.
In the following, we have to calculate the expectation values of
certain differential-position operators with respect
to the ground state wave functions.  
Since the norm of $u_{p} $ is one we obtain for the expectation value
of every differential operator of
first order with respect to $u_{p} $ a multiplicative factor $ \sqrt{p} $ in the result.
We get the same multiplicative factor $ \sqrt{p} $ for every position
operator in the differential-position operator. Thus, we can easily read off  
the leading $ p $ dependence  of the average of
a differential-position operator with respect to a lowest Landau level
wave function $u_{p} $.
Similarly, one can calculate the result of the average of a
differential-position operator over the many particle
ground state wave function $ u_{0,\vec{p}} $.

In order to calculate the asymptotic behaviour 
of the Green's function (\ref{1515}), we have organized this section
as follows:      
In subsection A, we will calculate the asymptotics of the Green's function 
$ G(\vec{r},\tau) $ for $ \vec{r}=0 $ and $ \tau>0 $. In subsection B
we will extend the calculation to $ \vec{r} \not= 0 $. In subsection C, we 
take the average of the asymptotic expressions of subsections A, B with
respect to the ground states of filling
fraction $ \nu=1/\tilde{\phi} $. The asymptotics of the Green's function 
$ G(\vec{r},\tau) $ for $ \tau<0 $ will be calculated in subsection D.
In subsection E, we consider the Chern-Simons Green's function
by taking into account the Coulomb interaction between the electrons.

\subsection{The calculation of the asymptotic behaviour
  of the Chern-Simons Green's function 
\boldmath  $ G(0,\tau)_{\tau>0}$\unboldmath}

In this subsection, we calculate the asymptotics of the
Chern-Simons Green's function 
$ G(0,\tau)_{\tau>0}$ for $ \tau>0 $.  
In the following, we apply the inverse of the Chern-Simons
transformation to the expression (\ref{1515}). The result will be
represented in the one particle basis. We get 
\begin{eqnarray} 
& & G(\vec{r},t;\vec{r}\,',t')=
\frac{e^{((t-t')-\beta)(\frac{1}{2}-\mu)N}}{Z} \eqlabel{1520} \\
& & \times \sum\limits_{u_{0,\vec{k}}\in \nu_{1/\tilde{\phi}}}\bigg|A\bigg[ u_{0,\vec{k}}(\vec{r}_1,..,\vec{r}_{N})\,
\delta_{\vec{r}}(\vec{r}_{N+1}
) \bigg] \bigg| \nonumber \\
& & 
\exp\left[-i \tilde{\phi}\left(\sum\limits_{i=1}^{N+1}
\alpha(\vec{r}_i-\vec{r})-\alpha(\vec{r}_i-\vec{r}\,')\right)\right]
  \nonumber\\
& & \bigg|  A\bigg[\exp\big[-(t-t')
\big(H_{N+1}+\overline{H}_{N+1}\big)\big] \nonumber \\
& & \hspace{3cm} 
u_{0,\vec{k}}
(\vec{r}_1,..,\vec{r}_{N})\,\delta_{\vec{r}\,'}(\vec{r}_{N+1})\bigg] \bigg>. \nonumber 
\end{eqnarray}
The Hamilton operators $H_{N+1}$, $ \overline{H}_{N+1} $ are given by 
\begin{eqnarray}
 & & H_{N+1}  =  
\frac{1}{2}\left[-i\vec{\nabla}_{N+1}+\vec{A}(\vec{r}_{N+1})+\tilde{\phi} 
\sum_{i=1}^N \vec{f}(\vec{r}_{N+1}-\vec{r}_i)\right]^2 ,\nonumber  \\ 
& & \overline{H}_{N+1}= 
\sum_{i=1}^{N}\frac{1}{2}\left[-i\vec{\nabla}_{i}+\vec{A}(\vec{r}_{i})+\tilde{\phi} 
\vec{f}(\vec{r}_i-\vec{r}_{N+1})\right]^2 . \eqlabel{1530}
\end{eqnarray}
Here $ \beta=1/(k_B T) $ where $ k_B $ is the Boltzmann constant. $ Z $  is the
partition function of the Hamiltonian $ H_e $.  
$ A $  is the antisymmetrization operator. $ \nu_{1/\tilde{\phi}} $ are the
ground states of the $ \nu=1/\tilde{\phi} $ system.
We split  $ H_{N+1}+\overline{H}_{N+1} $ in its components    
$ H_{N+1}+\overline{H}_{N+1}=H_0+H_1+H_2+H_3+H_{4,1}+H_{4,2} $
by
\begin{eqnarray}
& & H_{0,1}  =
\sum_{i=1}^N\frac{1}{2}\left[-i\vec{\nabla}_{i}+\vec{A}(\vec{r}_i)\right]^2
-\mu \;, \eqlabel{1540} \\
& & H_{0,2}=\frac{1}{2}\left[-i\vec{\nabla}_{N+1}+\vec{A}(\vec{r}_{N+1})\right]^2
\nts{3} -\mu ,\eqlabel{1550}\\
& & H_1 =  \sum_{i=1}^{N} \tilde{\phi}^2 \vec{f} 
(\vec{r}_i-\vec{r}_{N+1})^2\;, \eqlabel{1570}\\
& & H_2  =  \frac{1}{2} \sum_{i \not= j=1}^{N} \tilde{\phi}^2 \vec{f} 
(\vec{r}_i-\vec{r}_{N+1}) \vec{f}(\vec{r}_j-\vec{r}_{N+1}),\eqlabel{1565} \\
& & H_3  =  \sum_{i=1}^{N} \vec{A}(\vec{r}_i) \tilde{\phi} \vec{f}
(\vec{r}_i-\vec{r}_{N+1})-\vec{A}(\vec{r}_{N+1}) \tilde{\phi} 
\vec{f}(\vec{r_i}-\vec{r}_{N+1}) \nonumber \\
& & = -N\frac{\tilde{\phi}}{2}
\;,\eqlabel{1560} \\
& & H_{4,1}  =  \sum_{i=1}^{N} \tilde{\phi} \vec{f}
(\vec{r}_i-\vec{r}_{N+1})\frac{\vec{\nabla}_i}{i} , \eqlabel{1575}\\
& & H_{4,2}=  -\sum_{i=1}^{N} \tilde{\phi} \vec{f}
(\vec{r}_i-\vec{r}_{N+1})\frac{\vec{\nabla}_{N+1}}{i}\;. \eqlabel{1580}
\end{eqnarray}
In (\ref{1560}), we inserted the vector potential $ \vec{A} $ in
the symmetric gauge $
\vec{A}(\vec{r})=\frac{1}{2}{y \choose -x} $. Since we have the translation
invariance \cite{hlr} of the Chern-Simons Green's function at the
filling fraction
$ \nu=1/ \tilde{\phi} $ it is possible to fix $ \vec{r}=0 $. 
To calculate the asymptotic behaviour
of the Chern-Simons Green's function, we use the
cumulant theorem 
\begin{equation}
\left< e^{A} \right>=e^{<A>_c+\frac{1}{2}<A^2>_c+\frac{1}{3!}<A^3>_c+...}\;.\eqlabel{1590}
\end{equation}
The 'connected' expectation values are defined by 
$  \langle A \rangle_c =  \langle A \rangle
\;,\,
\langle A^2 \rangle_c  =  \langle\left(A-\langle A \rangle \right)^2 \rangle
\;,\,\langle A^3 \rangle_c
=  \langle\left(A-\langle A\rangle \right)^3 \rangle\;, \ldots   $

We now split the Chern-Simons Green's functions in two parts 
$ G(\vec{r},t,\vec{r}\,',t')=G^1(\vec{r},t,\vec{r}\,',t')
+G^2(\vec{r},t,\vec{r}\,',t') $. $ G^1(\vec{r},t,\vec{r}\,',t') $ is the
relevant term for $ \vec{r} \to \vec{r}\,' $ and has the form 
$ h(t-t',\vec{\nabla}_{r'})
\exp[-(t-t')
1/2[-i\vec{\nabla}_{r'}+\vec{A}(\vec{r}\,')]^2]
\delta(\vec{r}-\vec{r}\,') $.
$ h(t-t',\vec{\nabla}_{r'}) $  is a power series in the derivations in
$ \vec{r}\,' $ with prefactors in  $ (t-t') $.
$ G^2(\vec{r},t,\vec{r}\,',t') $ is a regular function (not of a
distributional form).
In the following, we  will only calculate the constant coefficient of the
power series $ h $ in $\vec{\nabla}_{r'} $. 
We will see in subsection D that $ G $ consists only on
$ G^2 $ for the time ordering $t-t'<0 $.
Since the ground state energy can be calculated from the Green's function 
with this ordering \cite{ne1}, we are especially  interested in that case.
A second reason for calculating only the constant coefficient of the power
series $h(t-t',\vec{\nabla}_{r'}) $ is that we are primary interested in
a comparison of an approximative Green's function (see section III) with the
exact Green's function for small $ |t-t'| $.
In a perturbational calculation of physical quantities one
commonly has to calculate integrals of a function of Green's functions.
The main contribution of the Green's functions
to these integrals is given for small $ |t-t'| $
(the Green's function tends to zero for large $ |t-t'| $). 
Thus the approximated Green's function should be rather exact in this range.
We will see at the end of this subsection that the higher order powers of
$ h $ in $ \vec{\nabla}_{r'} $ contribute only  on order $ O(|t-t'|^2) $.

In this subsection, we will calculate
$ G^1(\vec{r},t,\vec{r}\,',t') $.
With the the help of a binomial expansion of the products  as well as the
form of the normalized lowest Landau level  wave functions (\ref{1518}), we
get   
\begin{eqnarray}  
 & & \big<u_p\big|
      \left[-1+\left(2\vec{f}(\vec{r})\frac{\vec{\nabla}_r}{i}\right)^2\right]^{n'}   
 \left[-1+2\vec{f}(\vec{r})\frac{\vec{\nabla}_r}{i}\right]^n  \eqlabel{1620} \\
& & 
 \times \left[2 \vec{f}(\vec{r})\frac{\vec{\nabla}_r}{i}\right]^l
 \frac{1}{r^{m}}\left[\frac{\vec{r}}{r}\vec{\nabla}_r\right]^{m}
\big|u_p\big> \nonumber \\
&= & \;\delta_{m,0}\, \delta_{n,0} \,
\delta_{n',0} \nonumber \\ 
& + &\frac{1}{p}\;\Bigg[ -\delta_{m,2}\; \frac{1}{4}
\delta_{n,0}\,\delta_{n',0} \nonumber \\
& & 
+\delta_{m,1} 
\left\{\frac{1}{2}\delta_{n',0}\,
\delta_{n,1}+\delta_{n',1}\,\delta_{n,0}+\frac{1}{2}\, l
\delta_{n,0}\, \delta_{n',0}\right\} \nonumber \\
& &+\delta_{m,0}\bigg\{\delta_{n',2}\, 4 \, \delta_{n,0} +\delta_{n',1}\left(2\delta_{n,1}+(2l
    +1)\delta_{n,0}\right) \nonumber \\
& & +\delta_{n',0}\left(\delta_{n,2}+l \,\delta_{n,1}+
    \frac{(l^2-l)}{2}\delta_{n,0}\right)\bigg\}\Bigg]+ 
    O\left(\frac{1}{p^2}\right)\;. \nonumber 
\end{eqnarray}
This relation is useful for getting the cumulant expectation values
(\ref{1590}) in an easy way. 
For calculating the average values in (\ref{1520})
we consider in the following only terms behaving like $\sum 1/p $
(this sum diverges for $ A \to \infty $). Terms of the form
$ \sum O(1/p^2) $ are 
convergent.
At first, we neglect $ H_0 $ in $H_{N+1}+\overline{H}_{N+1}$.
So we have to  calculate the connected average values of the form
$ \langle|(H_1+H_2+H_3+H_{4,1}+H_{4,2})^n |\rangle_c $.
By neglecting finite terms of the form  
$ \sum O(1/p^2) $ we get with the
help of equation (\ref{1620}) the following diverging terms:
\begin{eqnarray}
& & \left<u_{0,\vec{k}}\left|H_2\right|u_{0,\vec{k}}\right>_c
  = -\frac{1}{4}\tilde{\phi}^2 \nts{6}
  \sum\limits_{u_p,u_{p+1}\in u_{0,\vec{k}}} \nts{2}
\frac{1}{p}  \;, \nonumber  \\ 
& & \left<u_{0,\vec{k}}\left|H_1\right|u_{0,\vec{k}}\right>_c
  =  \frac{1}{2}\tilde{\phi}^2 \nts{4} \sum_{u_p \in u_{0,\vec{k}}}\nts{1}
\frac{1}{p}\;,  \eqlabel{1660} \\
& & \left<u_{0,\vec{k}}\left|(H_3+H_{4,1})^n\right|u_{0,\vec{k}}
\right>_c
 =  \delta_{n,2} \;
 \frac{1}{4}\tilde{\phi}^2 \nts{4} \sum_{u_p \in u_{0,\vec{k}}}\nts{1}
 \frac{1}{p} \,.\nonumber 
\end{eqnarray}
We see from the first term of this  equation that we have also summations 
over nearest neighbour $p$.  
Now we take into account the operator   
$ H_0 $ in the calculation of the connected average values.
For doing this, we will use the Campbell-Hausdorff formula 
\begin{equation}
e^C=e^B \cdot e^A  \;,\eqlabel{1680}
\end{equation}
with 
\begin{eqnarray}
B & = & (C-A)+\frac{1}{2}[A,C]+\frac{1}{12}[C+A,[A,C]] \eqlabel{1690} \\
 & & 
-\frac{1}{24}[A,[C,[A,C]]]+O([,[,[,[,[]]]]]) \;. \nonumber 
\end{eqnarray}
We now fix $ C=-(H_{N+1}+\overline{H_{N+1}})\tau $ and  $ A=-H_0 \tau  $
in equation (\ref{1680}). 
With the help of the  Campbell-Hausdorff formula (\ref{1680}) and the
cumulant theorem (\ref{1590}) one can discuss the connected average values in
$ B $. This results in a power series in $ \tau=(t-t') $.
We will show later on in this subsection that this power series
will not terminate.  
Because of the enormous effort of calculation, we will limit ourselve
to the concrete
calculation of the coefficients up to the order  $ \tau^3 $. 
It will be the aim of a later publication to make predictions of these
coefficients for higher powers of $ \tau $ and further on the behaviour of
this power series for large $ \tau $.

The relevant cumulant expectation values for calculating the coefficients
up to the order $ \tau^3 $ are shown in appendix A.
With the help of these cumulant expectation values and the 
Campbell-Hausdorff formula 
(\ref{1680}), we get for the asymptotics of the 
Green's function for $ \tau>0$: 
\begin{eqnarray}
& & G^1(\vec{r},\tau)_{\tau>0} =  
 \nts{5}  \sum\limits_{u_{0,k} \in \nu_{1/\tilde{\phi}}} \nts{6} \exp \nts{1} \bigg[
\tilde{\phi}^2 \bigg\{\sum\limits_{u_p\in u_{0,k}}
\nts{1}\frac{1}{p} \Bigg(-\frac{1}{2}\tau
+\frac{1}{8} \tau^2  \nonumber \\ 
& & \qquad \qquad \qquad -\frac{1}{48}\tau^3+ O(\tau^4)\Bigg)  
 +O\left(\frac{1}{p^2}\right) \eqlabel{1700}  \\
& & + \nts{5}\sum\limits_{u_p,u_{p+1}\in u_{0,k}}\nts{3}\frac{1}{p}  
\left( \nts{1}\frac{1}{4}\tau \nts{1}- \nts{1}\frac{1}{48} \tau^3
  \nts{1}+ \nts{1} O(\tau^4)\right)+ \nts{1}
  O\left(\frac{1}{p^2}\right)\nts{2} \bigg\}\bigg]   
\frac{1}{\sum\limits_{u_{0,k}\in \nu_{1/\tilde{\phi}}}\nts{5} 1} \nonumber \\
& & \times \exp\left[-\tau
\frac{1}{2}\left[-i\vec{\nabla}_{r}+\vec{A}(\vec{r})\right]^2
\right]\delta(\vec{r}) 
.\nonumber 
\end{eqnarray}
We see from equation  (\ref{1700}) that the cumulant expectation values 
up to the order $\tau^3 $  behave like 
$\sum_{u_p \in u_{0,\vec{k}}} O(1/p) + \sum_{u_p,u_{p+1}\in u_{0,\vec{k}}} O(1/p) $.
In the following, we will show that equation (\ref{1700}) is correct for all
orders of $ \tau $. For doing this, we consider the operator $ H_{0,1} $
in polar coordinates:  
\begin{equation} 
 H_{0,1}=\sum\limits_{i=1}^{N}
 \frac{1}{2}\left[-\frac{1}{r_i}\,\partial_{r_i}-\partial_{r_i}^2
   +\left(-\frac{1}{r_i^2}\partial_{\phi_i}^2+i\partial_{\phi_i}
     +\frac{r_i^2}{4}\right)\right] \;\eqlabel{1705}.
 \end{equation}
The expectation value of the commutator of the
operator $ (1/r_i) \partial_{r_i} $ with the last three
summands of (\ref{1705}) 
results in the first factor on the left hand side in (\ref{1620}).  
Furthermore, one sees  that the second term of the product in (\ref{1620})
corresponds to $ (H_3+H_{4,1}) $. Thus, we get from (\ref{1620}) and the
form of $ H_{0,1} $ that by neglecting $ H_{0,2} $ and  $ H_{4,2} $
in $ H_{N+1}+\overline{H}_{N+1} $ the cumulant expectation values of the 
Campbell-Hausdorff terms behave like $\sum_{u_p \in u_{0,\vec{k}}}O(1/p)
+\sum_{u_p,u_{p+1}\in u_{0,k}} O(1/p) $.
In the following, we consider the cumulant expectation values of $ B $
(\ref{1690}) containing at least one of the terms $ H_{0,2} $, $H_{4,2} $. 
It is clear by  having more than two terms $ H_{4,2} $ in the cumulant
expectation value, we get as a result a convergent sum $ \sum O(1/p^2) $.
The number of operators
$ H_{0,2} $ is arbitrary.  Thus, we can limit our consideration to one or two
operators $ H_{4,2} $ in the cumulant expectation value. Because of the
simple structure of $ H_{N+1}+\overline{H}_{N+1} $ 
we get as a result of the cumulant expectation value
a term of the form $ \sum_{u_p\in u_{0,k}}(a_1/p+O(1/p^2))
+\sum_{u_p,u_{p+1}\in u_{0,k}}(a_2/p+O(1/p^2)) $. Here $ a_1 $, $a_2 $ are
real coefficients. 
Summarizing, we see that the exponent of the Green's functions
$ G^1(0,\tau)_{\tau>0} $
behaves like (\ref{1700}). 

\subsection {The asymptotic behaviour of the Chern-Simons Green's function  
 \boldmath $ G(\vec{r},\tau)_{\tau>0} $ \unboldmath
 for \boldmath $\vec{r} \not=0 $ \unboldmath }

In this subsection, we calculate $ G(\vec{r},\tau)_{\tau>0} $ for $ \tau>0 $ and
$\vec{r} \not=0 $. 
Because of the additional phase factor, we get
from equation (\ref{1520}) that the calculation of the asymptotic behaviour
of the
Green's function for $\vec{r} \not=0 $  is substantially more
difficult than for $ \vec{r}=0 $. To handle this phase factor 
we define
\begin{eqnarray}
& & \exp\left[M[N']\right]:=\exp\left[i\tilde{\phi}
  \left(\sum\limits_{i=1}^{N'}\alpha(
\vec{r}_i-\vec{r}\,')-\alpha(\vec{r}_i-\vec{r})\right)\right] \nonumber
 \\
& & =  \exp\bigg[i \tilde{\phi}
  \bigg(\sum\limits_{i=1}^{N'}\vec{f}(\vec{r}_i-\vec{r}\,')\cdot
(\vec{r}-\vec{r}\,')  \eqlabel{1780} \\
& & +(\vec{f}(\vec{r}_i-\vec{r}\,')
\cdot \vec{e}_x)\, (\vec{e}_y \cdot \vec{f}(\vec{r}_i-\vec{r}\,'))
  \nonumber  \\
& &\times \left(((\vec{r}-\vec{r}\,')
\cdot \vec{e}_x)^2  - (\vec{e}_y \cdot
(\vec{r}-\vec{r}\,'))^2\right)  \nts{1}  +\nts{1}O\left(\frac{1}{|\vec{r}_i-\vec{r}\,'|^3}
\right)\nts{1} 
\bigg)\bigg].\nonumber 
\end{eqnarray}
with $ 0 \le N' \le N $.   
$\vec{e}_x$ and $\vec{e}_y $, respectively, are unit vectors in 
the $x$- and $y$-direction, respectively.
When  calculating the Green's function (\ref{1520}), we get one term which
is proportional to $ \delta(\vec{r}-\vec{r}\,') $. This term was calculated
in the last subsection. Additionally, we get 

\begin{eqnarray}
 \lefteqn{G^2(\vec{r},t;\vec{r}\,',t')=
 -\frac{e^{((t-t')-\beta)(\frac{1}{2}-\mu)N}}{Z}}\eqlabel{1720} \\
 & &
 \nts{1} \times \nts{3} \sum\limits_{u_{0,k} \in \nu_{1/\tilde{\phi}}} 
 \sum\limits_{u_L, u_R \in u_{0,\vec{k}}} \nts{3} \left<V_L[u_{0,\vec{k}}] |
  e^{M[N-1]} e^{-(t-t') H_2}
 |V_R[u_{0,\vec{k}}] \right> \nonumber \\
& &  \times  
\left(e^{-(t-t')
    \frac{1}{2}\left[-i\vec{\nabla}_{r'}+\vec{A}(\vec{r}\,')-\tilde{\phi} 
\vec{f}(\vec{r}\,'-\vec{r})\right]^2} 
u_L(\vec{r}\,')\right)   \nonumber \\
& & 
\times  \left(e^{-(t-t')\frac{1}{2}\left[-i\vec{\nabla}_{r}+\vec{A}(\vec{r})+\tilde{\phi} 
\vec{f}(\vec{r}-\vec{r}\,')\right]^2}  u_R(\vec{r})\right) \nonumber 
\end{eqnarray}
with
\begin{equation}
H_2= 
\sum_{i=1}^{N-1}\frac{1}{2}\left[-i\vec{\nabla}_{r_i}+
  \vec{A}(\vec{r}_{i})+\tilde{\phi} 
\vec{f}(\vec{r}_i-\vec{r}\,')\right]^2 \eqlabel{1730}
\end{equation}
$ V_p[u_{0,\vec{k}}] $ is defined by
\begin{equation}\eqlabel{1740}
V_p[u_{0,\vec{k}}]= u_{0,\vec{k}'}
\sum\limits_{i=1}^{N} (-1)^{i+1} \, 
\delta_{p,k_i} \;.
\end{equation} 
$ \vec{k}' \in {\mathbb R}^{N-1}$ is up to
the entry which contains p given  by the vector $ \vec{k} $. 
In the following, we will transform (\ref{1720}) such that we can use the
results of the last subsection.
For doing this, we define the following operators: 
\begin{eqnarray}
& & O_{\gamma} :=  \gamma \Big(e^{-\tau\frac{1}{2}\left[-i\vec{\nabla}_{r'}+\vec{A}(\vec{r}\,')-\tilde{\phi} 
\vec{f}(\vec{r}\,'-\vec{r})\right]^2} \eqlabel{1750} \nonumber \\
& & \qquad \qquad \times \sum\limits_{i=1}^N 
P\left[\vec{r}\,'-\vec{r}\right](\vec{r}_i)\, \delta(\vec{r}_i-\vec{r})
\Big)\;, \nonumber \\
& & P\left[\vec{r}\,'-\vec{r}\right](\vec{r}_i) :=  1+(\vec{r}\,'-\vec{r})
\cdot\vec{\nabla}_i+\frac{1}{2} \; \ldots  \;,  \eqlabel{1760}  \\
& & H_M(\gamma,\tau) :=  
\overline{H}_{N+1} 
-\frac{1}{\tau}O_\gamma
-\frac{1}{\tau} M[N] \;.\eqlabel{1770}  
\end{eqnarray}
The operator $ P[\vec{r}\,'-\vec{r}](r_i) $ is a translation operator
which yields by applying it to a function  $ f(\vec{r}_i) $ the
translated function
$ f(\vec{r}_i+\vec{r}\,'-\vec{r}) $. 
Let us suppose that $ R $  is a power expansion of the operators $ O_\gamma$, $ M[N´]$
and further on other operators. 
Then we define the normal ordering of the operator $ : R : $
such that all operators $ O_\gamma $  are at the  left of the monomials
and the following operators are the $ M[N] $. 
With the help of the function 
\begin{eqnarray}
 F_M[\gamma,\tau] & := & 
\exp\bigg[-\tau\left\langle\left|:H_M(\gamma,\tau):\right|\right\rangle_c
\eqlabel{1782} \\
& &
+\frac{1}{2}\tau^2\;\left\langle\left|:H_M(\gamma,\tau)^2:\right|\right\rangle_c+...\bigg]
\;, \nonumber 
\end{eqnarray}
we get 
\begin{eqnarray}
G^2(\vec{r},t;\vec{r}\,',t') & = & 
 -\frac{e^{(t-t'-\beta)(\frac{1}{2}-\mu)N}}{Z}
 \sum\limits_{u_{0,k} \in \nu_{1/\tilde{\phi}}}
\, e^{-i \tilde{\phi}\, \alpha(\vec{r}-\vec{r}\,')}  \nonumber \\ 
& &  \times \frac{\partial}{\partial\gamma} F_M[\gamma,t-t'] \Bigg|_{\gamma=0}
 \;. \eqlabel{1785} 
\end{eqnarray}
The derivation with respect to $ \gamma $
of $F_M[\gamma,t-t'] $ at  $\gamma=0 $ is given by    
\begin{eqnarray}
 & & \frac{\partial}{\partial\gamma}
\left.F_M[\gamma,\tau]\right|_{\gamma=0}= F_M[0,\tau] 
 \frac{\partial}{\partial\gamma}\bigg(
\eqlabel{1790}\\ 
& & \left. -\tau\left\langle\left
|:H_M(\gamma,\tau):\right|\right\rangle_c \nts{2}+\nts{2} \frac{1}{2}\tau^2\left\langle\left|:H_M(\gamma,\tau)^2:\right|
              \right\rangle_c+..\bigg)\right|_{\gamma=0} \,. \nonumber 
\end{eqnarray}
We now split $ \overline{H}_{N+1} $
(\ref{1770}) in its components 
\begin{eqnarray}
& & \overline{H}_{N+1} \nts{1}= \nts{1}
\sum_{i=1}^N \nts{1} \left(\frac{1}{2} \nts{1}\left[-i\vec{\nabla}_{i}
  \nts{1} + \nts{1}\vec{A}(\vec{r}_i)\right]^2
\nts{4}-\mu \right)
\nts{1} + \nts{1} \frac{1}{2} \sum_{i=1}^{N} \tilde{\phi}^2 \vec{f} 
(\vec{r}_i-\vec{r}\,')^2  \nonumber  \\
& &+\left(-N\frac{\tilde{\phi}}{2}+ \sum_{i=1}^{N} \tilde{\phi} \vec{f}
(\vec{r}_i-\vec{r}\,')\frac{\vec{\nabla}_i}{i} \right) \,.
                     \eqlabel{1792}
\end{eqnarray}
As in the last subsection, we use the Campbell-Hausdorff formula
(\ref{1680}) to isolate  
$ H_{0,1} $ from
$ \overline{H}_{N+1} $ and
$ 1/2[-i\vec{\nabla}_{r'}+\vec{A}(\vec{r}\,')]^2 $ from
$ 1/2 [-i\vec{\nabla}_{r'}+\vec{A}(\vec{r}\,')-\tilde{\phi} 
\vec{f}(\vec{r}\,'-\vec{r})]^2 $. We now set $ \vec{r}\,'=0$. 
Then we can use the commutator analysis of appendix A and the results of the
last subsection.
First, we get
that the second term of the product in (\ref{1790}) does not contain any terms of
the form $ \sum 1/p $
(under the consideration that the $l $-th derivative of 
$ u_{p}(\vec{r}\,') $ with respect to $  \vec{r}\,' $ is at the
position $ \vec{r}\,'=0 $ zero for all $ p>l $). 
The first term in the product (\ref{1790}) is given by 
\begin{eqnarray}
& & G^2(\vec{r},\tau)_{\tau>0}=   \eqlabel{1800} \\ 
&&  \sum\limits_{u_{0,k} \in \nu_{1/2}} \nts{6} \exp \nts{1} \bigg[
\tilde{\phi}^2 \bigg\{\sum\limits_{u_p\in u_{0,k}}
\nts{1}\frac{1}{p} \Big(-\frac{1}{4} r^2-\frac{1}{4} \tau
+\frac{1}{8}
\tau^2 -\frac{1}{24} \tau^3   \nonumber \\
& & 
+ O(\tau^4)\Big)+O\left(\frac{1}{p^2}\right)+\nts{6}\sum\limits_{u_p,u_{p+1}\in u_{0,k}}
\nts{6}\frac{1}{p} \frac{1}{4} r^2 \nts{1} + \nts{1} O\left(\frac{1}{p^2}\right)   
 \Bigg\} \nts{2} \Bigg]. \nonumber 
\end{eqnarray}
We see from $ G^2(\vec{r},\tau)_{\tau>0} $ that the
$ \sum 1/p $ coefficients of the higher order $ \tau $ terms
($\tau>3 $) have no $ \vec{r} $ dependence. Furthermore, one gets from
the form of $ \overline{H}_{N+1} $ and the relation (\ref{1620}) 
that the higher order $ \tau $ terms have not to be zero.

\subsection{Averaging of the Green's function with respect to
  the ground states}

From the equations  (\ref{1700}) and  (\ref{1800}), we have to calculate 
an expression of the following form 

\begin{eqnarray}
& &
 G(\vec{r},\tau) =\sum\limits_{u_{0,\vec{k}} \in \nu_{1/\tilde{\phi}}}\nts{2}\exp\bigg[
 \sum\limits_{u_p\in u_{0,\vec{k}}} 
            (\tilde{\phi}^2 f_1(r,\tau))
            \frac{1}{p}+O\left(\frac{1}{p^2}\right) 
 \nonumber \\
 & &
       + \nts{4} \sum\limits_{u_{p},u_{p+1}\in u_{0,\vec{k}}}
         (\tilde{\phi}^2 f_2(r,\tau))
            \frac{1}{p} +O\left(\frac{1}{p^2}\right)\bigg]
          \frac{1}{\sum\limits_{u_{0,\vec{k}}\in \nu_{1/2}}\nts{4}1} \;.
        \eqlabel{1810} 
 \end{eqnarray}
$ f_1(r,\tau) $ and $f_2(r,\tau) $ are real functions of 
$ (r,\tau) $. \\
In the following, we will show that the asymptotic behaviour 
of the result is given by 
\begin{equation}\eqlabel{1820}
G(\vec{r},\tau)=\exp\left[-2\left(\tilde{\phi}f_1(r,\tau)+f_2(r,\tau)\right)\log(c)
  +O(1)\right]     \;.
\end{equation} 
Here $ c $ is proportional to $ 1/\sqrt{A} $. So we get   
$\log(c)=-\frac{1}{2}\log(N)+O(1)$. 
With the help of the expression 
\begin{eqnarray}
\lefteqn{E[u_{0,\vec{k}}]=} \eqlabel{1830} \\
& & \sum\limits_{u_p\in u_{0,\vec{k}}} 
            (\tilde{\phi}^2 f_1(r,\tau))\left
            (\frac{1}{p}-\frac{1}{N\tilde{\phi}}
            (\log(N)+{\cal C})\right)\nts{1}+\nts{1}O\left(\frac{1}{p^2}\right)\nonumber  \\
& &   +\nts{10}\sum\limits_{u_p, u_{p+1}\in u_{0,\vec{k}}}\nts{7} 
            (\tilde{\phi}^2 f_2(r,\tau))\left
            (\frac{1}{p}-\frac{1}{N\tilde{\phi}^2}
            (\log(N)+{\cal C})\right)\nts{1}+\nts{1}O\left(\frac{1}{p^2}\right), \nonumber     
\end{eqnarray}
the asymptotics (\ref{1820}) is correct if we show that the 
following expression $ K $ is finite for $ A \to \infty $: 

\begin{equation}
  K =\sum\limits_{u_{0,\vec{k}} \in \nu_{1/\tilde{\phi}}}\exp\left[
      E\left[u_{0,\vec{k}}\right]\right]
  \cdot \frac{1}{\sum\limits_{u_{0,\vec{k}}\in \tilde{\phi}} \nts{4} 1} \;.\eqlabel{1840} 
 \end{equation}
$ {\cal C} $ is the Euler number which is defined by 
$  \sum_{p=1}^{N} 1/p=\log(N)+{\cal C}+
O(1/N) $.
We will determine  $ K $ by the calculation of the moments $ M_n $  of the operator 
$ \exp[E(u_{0,\vec{k}})]$. The moments are defined by 
$  M_n=\sum_{u_{0,\vec{k}} \in \nu_{1/2}}(E[u_{0,\vec{k}}])^n
 \cdot 1/(\sum_{u_{0,\vec{k}}\in \nu_{1/2}} \nts{4}1)$. 
From this definition we obtain $ M_0=1 $.\\
For calculating  $ M_1 $, we use the following transformation
\begin{eqnarray}
& & \frac{1}{\nts{4}\sum\limits_{u_{0,\vec{k}}\in \nu_{1/\tilde{\phi}} }\nts{4} 1} 
\sum\limits_{u_{0,\vec{k}} \in \nu_{1/2}} \sum\limits_{u_p\in u_{0,\vec{k}}} \frac{1}{p}
 =  \frac{1}{\nts{4}\sum\limits_{u_{0,\vec{k}}\in \nu_{1/\tilde{\phi}}} \nts{4}1} 
 \sum\limits_{p=1}^N\,\frac{1}{p}\; {N-1 \choose 
  \frac{N}{\tilde{\phi}}-1} \nonumber   \\
& &  = \frac{1}{\tilde{\phi}}\left(\log(N)+{\cal C}+O\left(
 \frac{1}{N}\right)\right)  \;, \eqlabel{1870}  \\
& & \frac{1}{\nts{4}\sum\limits_{u_{0,\vec{k}}\in \nu_{1/\tilde{\phi}}}\nts{4} 1} 
\sum\limits_{u_{0,\vec{k}} \in \nu_{1/2}} 
\sum\limits_{u_p, u_{p+1}\in u_{0,\vec{k}}}
\frac{1}{p}
  = \frac{1}{\nts{4}\sum\limits_{u_{0,\vec{k}}\in \nu_{1/\tilde{\phi}}}\nts{4} 1} 
 \sum\limits_{p=1}^N \, \frac{1}{p} \; {N-2 \choose 
  \frac{N}{\tilde{\phi}}-2}\nonumber \\
 & &  =  \frac{1}{\tilde{\phi}^2}\left(\log(N)+{\cal C}+O\left(
 \frac{1}{N}\right)\right)  \;. \eqlabel{1875}
\end{eqnarray}
Thus, we get that  $ M_1 $ is finite. \\[0.2cm]
For calculating  the higher moments, we will show at first
that $ A_{(M_1+M_2)} $ is finite. $ A_{(M_1+M_2)} $ is defined by  
\begin{eqnarray}
& & A_{(M_1+M_2)} =
\lim\limits_{N \to \infty}\nts{4} 
\sum\limits_{u_{0,\vec{k}}\in \nu_{1/\tilde{\phi}}} \Bigg[
  \prod\limits_{i'=1}^{M_2} \sum\limits_{u_p\in u_{0,\vec{k}}}
            \frac{1}{p^{m'_i}}
      \eqlabel{1880} \\
 & &        \times
\prod\limits_{i=1}^{M_1}
\left(\sum\limits_{u_p\in u_{0,\vec{k}}} 
           \nts{4}  \left(\frac{1}{p}-\frac{1}{N\tilde{\phi}}
            (\log(N)+{\cal C})\right)^{n_i} \nts{4} \frac{1}{p^{m_i}}\right)
\cdot \frac{1}{\nts{4} \sum\limits_{u_{0,\vec{k}}\in \nu_{1/\tilde{\phi}}}\nts{4}1} \Bigg]   \;. \nonumber 
\end{eqnarray}
We are interested in $ A_{(M_1+M_2)} $ for the values 
$ n_i\ge 1 $, $ m_{i'}\ge 2$, $ m_i\ge 2 $ or $m_i=0 $.
To show the finiteness of $ A_{(M_1+M_2)} $ we carry out a proof by
induction. \\
Because of the equations (\ref{1870}), (\ref{1875})
and the finiteness of $ \sum_{u_p \in u_{0,\vec{k}}\in \nu_{1/2}} 1/p^n $
for $ n \ge 2 $ we get that  $ A_1 $ is finite. 
For  $ J>1 $ we can split $ A_J $ in the following two terms: 
\begin{eqnarray}
& &  A_{J}= \lim_{N \to \infty}\nts{2} 
\sum\limits_{u_{0,\vec{k}}\in \nu_{1/2}} \nts{2} 
\Bigg[\sum\limits_{u_{p_i},u_{p_{i'}}\in u_{0,\vec{k}}
    \atop p_i \not=p_{i'}}
\prod\limits_{i'=1}^{M_2} 
           \left( \frac{1}{p_{i'}^{m_{i'}}}\right)
\eqlabel{1910}\\
 & & 
     \prod\limits_{i=1}^{M_1}
            \left
            (\frac{1}{p_i}-\frac{1}{N\tilde{\phi}}
            (\log(N)+{\cal C})\right)^{n_i} \nts{2}
          \left(\frac{1}{p_i^{m_i}}\right) \nts{2}
          \cdot \frac{1}{\nts{4}
            \sum\limits_{u_{0,\vec{k}}\in \nu_{1/\tilde{\phi}}}\nts{4}1}
           \Bigg] +  A_{J-1} .\nonumber 
\end{eqnarray}
By  $ p_i \not=p_{i'} $ we mean that all $
p_i $, $ p_{i'} $   are different.
Due to the induction assumption, we only have to discuss the first term 
 $ A_{J,1} $ in $  A_{J} $.
With the help of the induction assumption we get for $ A_{J,1} $ as in
the calculation of $ M_1 $
\begin{eqnarray}
& & A_{J,1} =   
\frac{1}{\tilde{\phi}^{|M_1|+|M_2|}}\prod\limits_{i=1}^{M_1}
\prod\limits_{i'=1}^{M_2}\sum\limits_
{p_i,p_{i'}=1}^{N}
            \Bigg[\left(\frac{1}{p_{i'}^{m_{i'}}}\right)\eqlabel{1930} \\
& & 
\left(\frac{1}{p_i}-\frac{1}{N\tilde{\phi}}
            (\log(N)+{\cal C})\right)^{n_i} 
            \left(\frac{1}{p_i^{m_i}}\right)
           \Bigg]+\mbox{finite}\,.
         \nonumber 
       \end{eqnarray}
From this we see that $ A_{J,1} $ is finite. Thus, we get the finiteness
of $ A_{(M_1+M_2)} $ for $ N \to \infty $. 
\\[0.2cm]
Summarizing, we obtain the finiteness of  $ M_n $ for every $ n $ if
$ f_2(r,\tau)=0 $. It is easy to generalize the considerations above to the
case of $ f_2(r,\tau) \not=0 $. Thus the Chern-Simons Green's function
$ G(\vec{r},\tau)_{\tau>0} $ has the form of equation (\ref{1820}). 
With the help of (\ref{1700}) and  (\ref{1800}) we get for the 
asymptotic behaviour of the Green's function
$ G(\vec{r},\tau)_{\tau>0}=G^1(\vec{r},\tau)_{\tau>0}+
G^2(\vec{r},\tau)_{\tau>0} $ 
\begin{eqnarray}
& &  G^1(\vec{r},\tau)_{\tau>0}  \sim   
\exp\nts{1} \Bigg[\nts{1} -2 \tilde{\phi} \log(c)\nts{1}\Bigg\{
    \nts{1}- \left(\frac{1}{2}-\frac{1}{4 \tilde{\phi}
        }\right)\frac{B}{m}\tau
    \eqlabel{1940} \\
 & & +\frac{1}{8}\left(\frac{B}{m}\tau\right)^2 -
 \left(\frac{1}{48}+\frac{1}{48 \tilde{\phi} }\right)
\left(\frac{B}{m}\tau\right)^3+O(\tau^4) \Bigg\}\Bigg] \nonumber \\
& & \times \exp\left[-\tau
\frac{1}{2m}\left[-i\vec{\nabla}_{r}+\vec{A}(\vec{r})\right]^2 \right]
\delta(\vec{r})  \nonumber 
\end{eqnarray}
and
\begin{eqnarray}
& & G^2(\vec{r},\tau)_{\tau>0}\sim\exp\nts{1} \Bigg[\nts{1} -2 \tilde{\phi} \log(c)\nts{1}\Bigg\{
    \nts{1} -  \left(\frac{1}{4}-\frac{1}{ 4 \tilde{ \phi}
        }\right) \frac{B}{m} r^2 \eqlabel{1945} \\
& & -\frac{1}{4} \, \frac{B}{m}\tau
+ \frac{1}{8}
\left(\frac{B}{m}\tau\right)^2
-\frac{1}{24} \left(\frac{B}{m}\tau \right)^3
+O(\tau^4)\nts{2} \Bigg\} \nts{2} \Bigg]. \nonumber 
\end{eqnarray}

\subsection{The asymptotic behaviour of the Chern-Simons Green's
function  \boldmath $ G(\vec{r},\tau)_{\tau<0} $ \unboldmath }

In this subsection, we calculate the Chern-Simons Green's function 
$ G(\vec{r},\tau)_{\tau<0} $ for $\tau<0 $. 
At first, we define a wave function originating from a $ \nu=1/\tilde{\phi} $
wave function in which a particle at position $ \vec{r} $ is annihilated.
This wave function is given by 
\begin{equation}\eqlabel{1950}
 u'_{0,\vec{k}}(\vec{r}_1,..\vec{r}_{N-1};\vec{r}):=
\sum\limits^{N}_{i=1}\int d\vec{r}_i\,
u'_{0,\vec{k}}\,(\vec{r}_1,..\vec{r}_{N})\,\delta(\vec{r}_i-\vec{r}) 
\end{equation}
(in the definition  of equation (\ref{1950}) we implicitly carried out
a renaming of the indices).
Similar to our calculation in subsection A we get for  $ t-t' <0 $ from the definition
(\ref{1515})   
\begin{eqnarray}
& & G(\vec{r},t;\vec{r}\,',t')=
  \frac{e^{((t'-t)-\beta)(\frac{1}{2}-\mu)N}}{Z}\eqlabel{1960} \\
& & \sum\limits_{u_{0,k}}\bigg<u'_{0,k}(\vec{r}_1,..,\vec{r}_{N-1};\vec{r})
\bigg|  \nonumber \\
& & \exp\left[-i\tilde{\phi}\left(\sum\limits_{i=1}^{N-1}
\alpha(\vec{r}_i-\vec{r})-\alpha(\vec{r}_i-\vec{r}\,')\right)
\right]
 \nonumber \\
& & 
\times \exp\left[-(t'-t)\overline{H}_N\right] \bigg|u'_{0,k}
(\vec{r}_1,..,\vec{r}_{N-1};\vec{r}\,')\bigg> \nonumber 
\end{eqnarray}
with the Hamilton operator 
\begin{equation}\eqlabel{1970}
\overline{H}_N
=\sum_{i=1}^{N-1}\frac{1}{2}\left[-i\nabla_{i}
+\vec{A}(\vec{r}_{i})+\tilde{\phi} \vec{f}(\vec{r}_i
-\vec{r}\,')\right]^2 \;.
\end{equation}
As in the last subsections, we split $ \overline{H}_N $ in its components 
\begin{eqnarray}
& & \overline{H}_N \nts{1}= \nts{2}
\sum_{i=1}^{N-1} \nts{2}\left( \frac{1}{2} \nts{1}\left[-i\vec{\nabla}_{i}
  \nts{1} + \nts{1}\vec{A}(\vec{r}_i)\right]^2
\nts{4}-\mu \right)
\nts{1} + \nts{1} \frac{1}{2} \nts{1}  \sum_{i=1}^{N-1} \tilde{\phi}^2 \vec{f} 
(\vec{r}_i-\vec{r}\,')^2  \nonumber  \\
& &+\left(-(N-1)\frac{\tilde{\phi}}{2}+ \sum_{i=1}^{N-1} \tilde{\phi} \vec{f}
(\vec{r}_i-\vec{r}\,')\frac{\vec{\nabla}_i}{i} \right) \;.
                     \eqlabel{1980}
\end{eqnarray}
Now we can use the commutator analysis of appendix
A and the results of the last subsections to get for the Chern-Simons Green's
function $ G(\vec{r},\tau)_{\tau<0} $ 
\begin{eqnarray}
& & G(\vec{r},\tau)_{\tau<0} \sim   
\exp\nts{1} \Bigg[\nts{1} -2 \tilde{\phi} \log(c)\nts{1}\Bigg\{
    \nts{1} -  \left(\frac{1}{4}-\frac{1}{ 4 \tilde{ \phi}
        }\right) \frac{B}{m} r^2 \eqlabel{2020} \\
& & +\frac{1}{4} \, \frac{B}{m}\tau
+\frac{1}{8}
\left(\frac{B}{m}\tau\right)^2
+\frac{1}{24} \left(\frac{B}{m}\tau \right)^3
+O(\tau^4)\nts{2} \Bigg\} \nts{2} \Bigg]. \nonumber 
\end{eqnarray}

\begin{figure}[t]
  \psfrag{G1}{}
  \psfrag{G2}{}
\centerline{\psfrag{xxx}
   {\smash{\raisebox{-0.05cm}{\footnotesize $ \frac{B}{m} |\tau|$}} }
   \psfrag{y}{\turnbox{180}{\footnotesize $ \nts{7}
       \frac{\log(G)}{(- 4 \log(c))}$}}  
 \epsfig{file=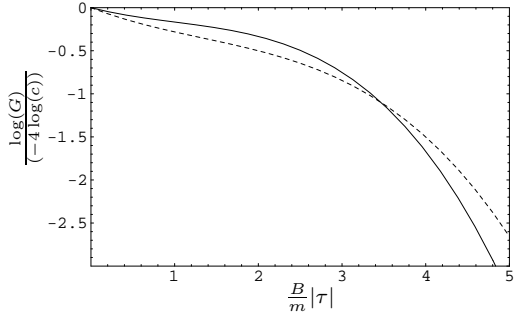,width=6.5cm}
  }
\hspace*{0.1cm}
 \caption{$ \log(G)/(- 4 \log(c)) $ for the $ \nu=1/2 $ system
   where $ G $ is either the first factor in (\ref{1940}) for $ \vec{r}=0 $
   (dashed line),
or  $G= G^2(0,|\tau|)_{\tau>0}=G(0,|\tau|)_{\tau<0} $ (solid
   line). \label{fig1}}\end{figure}

By taking the  limit $ A \to\infty $
we get from the  equations (\ref{1940}), (\ref{1945}) and (\ref{2020}) that 
the Chern-Simons Green's function $ G(\vec{r},\tau) $
vanishes exponentially for $ \tau>0 $ as well as for $ \tau<0 $.
This is illustrated  in figure \ref{fig1} where we show
the function $  \log(G)/(- 4 \log(c)) $ for the $ \nu=1/2 $ system.
$ G $ is either the first factor in
(\ref{1940})  for $ \vec{r}=0 $ or
$ G=G(0,|\tau|)^2_{\tau>0}=G(0,|\tau|)_{\tau<0} $
(\ref{1945}), (\ref{2020}).
The vanishing of the Green's function can be explained by the boundedness
of the Green's function (easily obtained from the definition (\ref{1515})).
It is an interesting question if the $ \log(c) $ behaviour is 
also true when considering also higher $ \tau $ terms in the concrete
calculation of
the Green's function. Since the number of summands of the
Campbell-Hausdorff formula grows very fast for large $ \tau $
we think there is
no easy answer to this question. As mentioned above, 
we are primary interested in this paper 
in the Green's function for small $ \tau $.
So instead of considering further this question, we
investigate the asymptotics of the Chern-Simons Green's function taking into
consideration the Coulomb interaction. 

\subsection{ The asymptotics of the Green's function taking into consideration
  the Coulomb interaction}

In the case of the Chern-Simons theory taking into consideration the Coulomb
interaction the ground state $ u_0 $ consists of a superposition of
slater determinants which consists of Landau wave functions of the lowest as
well as of higher Landau levels. 
In the following, we make the approximation that we consider in $ u_0 $ only
that part which consists of wave functions in the lowest Landau level.
This restriction on the wave function is a standard approximation to calculate
for example the ground state energy or the effective mass of the composite
fermions \cite{mo1}. The ground state wave function is given by    
$ u_0=\sum_{\vec{p}\in  {\mathbb N}_0^N}
 c_{\vec{p}} 
\; S\left[u_{p_1},u_{p_2},...\right]  $. The coefficients $ c_{\vec{p}} $
underlie the restriction 
$ \sum_{\vec{p}\in {\mathbb N}_0^N} 
\left|c_{\vec{p}}\right|^2=1  $. This is a result
of the normalization of $u_0 $.
To get insight into the Chern-Simons Green's function taking into
account the Coulomb interaction, we can transform the Green's function
similar to (\ref{1520}) and (\ref{1960}) with an additional sum over the
slater determinants of $ u_0 $. With the help of the Campbell-Hausdorff
formula we separate similarly to the last subsections the Hamilton
operator $ H_0 $ from the rest of the operators in
$ H_{N+1}+\overline{H}_{N+1}$, $  \overline{H}_{N+1} $ or$ \overline{H}_N $
($ H_0 $ also consists of the Coulomb part of the Hamilton operator).  
Suppose at first the restriction that the two Slater determinants which are the ingredients of the
average values in (\ref{1520}) and  (\ref{1960}) are not in agreement in
their Landau wave functions $u_{p_i} $ for $i =1,..,N_C $ with $ N_C \le N $.
Furthermore, suppose  that $ N-N_C $ is finite for $ A \to \infty $. 
Then, we get that the average value of $ e^B $ (\ref{1690})
with respect to these  two Slater
determinants behaves as $ \lesssim  \prod_{i=1,..,N_C} 1/p_i^{1/4}\lesssim
e^{-1/4 N_C} $ (the Coulomb operator in  a scalar product between
two Slater determinants of Landau wave functions $ u_{p_j} $ $j=1,..,4$
scales like $ \sim \mbox{Min}[\sqrt{1/p_j}] $. Here $ \mbox{Min}[\cdot] $
is the minimum of its argument). 
Since $ N_C \propto A  $ for $ A \to \infty $, we see that the average value 
with respect to the two Slater determinants vanishes exponentially for
$ A \to \infty $.

In the following, we consider the case of no  restrictions on the variables
$ N $,  $N_C$. Then we get from the equations (\ref{1590}), (\ref{1810})
that the asymptotic form  of the summands of the Green's function
behaves as 
$ \lesssim e^{-1/4 N_C} \cdot e^{-f(r,\tau,N,e^2)\log(N-N_C)} $.
$ f(r,\tau,N,e^2) $ is given by 
$f(r,\tau,N,e^2) = f(r,\tau) (\sum_{j=1..N} (1/p_j)-\sum_{i=1..N_C} (1/p_i))/
\log(N-N_C)  $ (we neglect the nearest neighbour sum in (\ref{1810})).
Furthermore, we have $0 \le \lim_{N \to \infty}
f(r,\tau,N,e^2) \le f(r,\tau)$ ($ \lim_{N \to \infty}
f(r,\tau,N,e^2) $ has not to be
convergent). Thus, we get that the maximum of the summands of the Green's
function is given by equal Slater determinants of which the average values
(\ref{1520}) and (\ref{1960}) are built. 
Summarizing, we get also in the case of the Chern-Simons theory
taking into account the Coulomb interaction the asymptotics  
(\ref{1940}), (\ref{1945}) and (\ref{2020}) of the Green's function, whereby
we have to replace $ \log(c) $ by $ V(e^2,c) \log(c) $
in these formulas. $ V(e^2,c) $ is a function of the Coulomb coupling
constant $ e^2 $ with $ 0 \le \lim_{c \to 0} V(e^2,c) \le \tilde{\phi} $ and
$ \lim_{c \to 0} V(0,c)=1 $. In principle $V(e^2,c)$ may have the
limit $ \lim_{c \to 0}  V(e^2,c)=0 $ for some $ e^2 $. Then we get
that the Green's function does not vanish exponentially for $ A \to \infty
$. Moreover $ \lim_{c \to 0}  V(e^2,c) $ does not have to
be continuous at $ e^2=0 $ for temperature $ T=0 $.
By considering the Green's function for $ T>0 $
(and limit the Hilbert space to the lowest Landau level) we get
for the asymptotics of the Green's function the equations
(\ref{1940}), (\ref{1945}) and  (\ref{2020}) with the replacement $ \log(c) $
by $ V(e^2,c)
\log(c) $ where $  V(e^2,c) $ is a function of the temperature.
For $ T>0 $ the limit $ \lim_{c \to 0}  V(e^2,c) $ has to be continuous 
at $e^2= 0$ with $ \lim_{e^2 \to 0 } \lim_{c \to 0}  V(e^2,c)=1 $.
Thus also in the case of the Chern-Simons theory taking into
consideration the Coulomb interaction, we obtain an asymptotic
behaviour of the Green's
function which vanishes for $ A \to \infty $ .

We should remark a consequence of the asymptotic behaviour
of the Green's function.
By inserting between
the creation and the annihilation operator in (\ref{1515}) a complete set of
eigen functions of the Hamiltonian $ H_{CS} $, we get   
that the overlap between  the wave functions
$ A[u_0(\vec{r}_1,..,\vec{r}_N),\delta(\vec{r}_{N+1}-\vec{r})] $ or
$ \sum^{N}_{i=1}\int d\vec{r}_i
u_0(\vec{r}_1,..,\vec{r}_{N}) \delta(\vec{r}_i-\vec{r}) $, respectively, 
and the eigen functions of the Hamilton operator $ H_{CS} $
with $ N+1 $ or $ N-1 $ particles, respectively, vanishes for $ A \to \infty $.

\section{The Green's functions in the Hartree-Fock approximation as
  well as in the RPA}

When calculating physical
quantities perturbatively the mean field Green's function of the
perturbation theory should have a similar form as the exact Green's function.
In the following, we compare the asymptotic behaviour of the Green's function
in the Hartree-Fock approximation as well as
in the RPA with the exact Green's function to make a step towards to this
perturbation theory . We will show that
in contrast to the RPA Green's function the Hartree-Fock Green's
function has a similar asymptotic behaviour as
the exact Green's function. Because in a perturbation theory
with auxiliary
fields one usually takes as a starting point
a Green's function \cite{ne1} which is determined self consistently, we will
calculate the self consistent
Hartree-Fock Green's function and compare
the calculated density of this Green's function 
with the exact particle density.
Since the Chern-Simons interaction is the reason for the asymptotic
vanishing form of
the Chern-Simons Green's for $ A \to \infty $, we neglect
at first the Coulomb interaction.

It is easy to calculate the Hartree-Fock approximation
of the Green's function of the Chern-Simons Hamiltonian (\ref{1050}) for
$ A \to \infty $. One gets
\begin{eqnarray}
& & \Sigma^{\tiny \mbox{HF}} (q,k_F) = \frac{\tilde{\phi}^2}{4}\frac{k_F^2}{m} 
\bigg[\log(4c^2) \eqlabel{2065}  \\
& & -\log\left(2\sqrt{(k_F^2+q^2+c^2)^2-4q^2k_F^2}
+2k_F^2+2(c^2-q^2)\right)\bigg]  \nonumber \\
& & -\frac{\tilde{\phi}^2}{4}\frac{k_F^2}{m}
\log\left(\frac{c}{k_F}\right)
+ \Sigma^{f
  } (q,k_F)  \;.  \nonumber  
\end{eqnarray} 
As in the last section  $ c $ is an impulse cut off in the infrared region.
To get $  \Sigma^{\tiny \mbox{HF}} (q,k_F) $ we calculated all Hartree-Fock Feynman
diagrams by inserting the interaction free Green's function
$ G(\vec{q},\omega)=-1/(i\omega-q^2/(2m)+\mu) $. For this
Green's function the fermi momentum  $ k_F $ is given by $ k_F=\sqrt{2m \mu} $.
By pairing the outer creation and annihilation operator in (\ref{1050})
we get an effective two particle operator which scales in the momentum
like $ \sim 1/q^2 $. The first summand in (\ref{2065}) represents
the exchange diagram of this vertex. The rest of the diagrams which 
diverge  for $ c \to 0 $ are given by the second summand. This term is 
given by a diagram of the Hartree form. The last term $ \Sigma^{f} $
is finite for
$ c \to 0 $ .  
The Hartree-Fock Green's function was also calculated by  Sitko and
Jacak \cite{sit2} before.
Carrying out the limit $ c \to 0 $ in (\ref{2065}) we get 
\begin{equation}
 \begin{array}{l} 
\Sigma^{\tiny \mbox{HF}} (q,k_F) = \\[0.3cm]  \eqlabel{2120}
\begin{array}{c c}
\nts{3}\left\{  
\begin{array}{l c}
\frac{\tilde{\phi}^2}{4}\frac{k_F^2}{m}\left[\log\left(\frac{c}{k_F}\right)
-\log\left(\frac{k_F^2-q^2}{k_F^2}\right)\right]+\Sigma^f(q,k_F) & \nts{3}q<k_F \\ 
-\frac{\tilde{\phi}^2}{4}\frac{k_F^2}{m}\left[\log\left(\frac{c}{k_F}\right)-\log\left(\frac{q^2-k_F^2}{q^2}
\right)\right]+\Sigma^f(q,k_F) & \nts{3} q>k_F \\
\Sigma^f(k_F,k_F)
& \nts{3} q=k_F .
\end{array}\right. 
\end{array}
\end{array} 
\end{equation}
Thus, we see that $ \Sigma^{\tiny \mbox{HF}} $ is finite on the
Fermi surface ($ q=k_F$). This is
accomplished by the changing of the sign of the $ \log(c) $ singularity on
the Fermi surface. We now calculate the Fourier (time) transform of the Hartree-Fock
Green's function (the Green's function which includes
the Hartree-Fock self energy $ \Sigma^{\tiny \mbox{HF}} $).
In the leading $ c $ order, we get for this Green's function   
\begin{eqnarray}
& & \frac{1}{2\pi} \int \; d\omega\; G^{\tiny \mbox{HF}}(q,\omega)\;
e^{-i\omega \tau} \eqlabel{2180} \\ 
& &
= -\;e^{[-\left( \frac{q^2}{2m}-\mu\right)-
\frac{\tilde{\phi}^2}{4}\frac{k_F^2}{m}\log(c/k_F)]\;\tau}n_F(q) \;
\Theta(-\tau) \nonumber \\ 
& & +\;e^{[-\left( \frac{q^2}{2m}-\mu\right)+
\frac{\tilde{\phi}^2}{4}\frac{k_F^2}{m}\log(c/k_F)]\;\tau}(1-n_F(q)) \;
\Theta(\tau) \;.\nonumber 
\end{eqnarray}
Here $ n_F(q) $ is the fermi factor
$ n_F(q)=1/(\exp[\beta(q^2/(2m)-\mu)]+1)$. 
When carrying out the Fourier transformation
with respect to $ \vec{q} $  and
comparing the result with the asymptotics  (\ref{1940}), (\ref{1945}) and
(\ref{2020})
of the exact Chern-Simons Green's function we obtain that the two asymptotics
are in accordance for small $ \tau $ and $ \vec{r}=0 $.  
Furthermore, we see that the prefactor of the $ \log(c/k_F) $ term in 
$ G^{\tiny \mbox{HF}}(q,\omega) $ is equal to the exact
Green's function.

Next, we calculate the Chern-Simons Green's function in RPA.
For doing this we use the path integral in \cite{di1} which includes
the bosonic Chern-Simons fields.
This path integral correspondence to the path integral of HLR
\cite{hlr} up to one additive term in the action. This additive term
is necessary to reproduce the correct ordering of the operators  \cite{di1} in
the Chern-Simons Hamiltonian (\ref{1050}).
In \cite{di1}, we calculated
the grand canonical
potential $ \Omega_{\tiny \mbox{RPA}} $ from this path integral
in RPA. In the following,
we will calculate the RPA
self energy through $ \Sigma^{\tiny \mbox{RPA}}=
\delta \Omega_{\tiny \mbox{RPA}}/(\delta G) $. After some calculation,
$ \Sigma^{\tiny \mbox{RPA}}$ is given
in the leading $ c $ order by 
\begin{eqnarray} 
\lefteqn{\Sigma^{\tiny \mbox{RPA}}(q,\omega)=} \eqlabel{9560}\\
& & \tilde{\phi}^2 \frac{\mu}{2}
\log\left(\frac{c}{k_F}\right)
\frac {\left(i \omega-\frac{q^2}{2m}+\mu\right)}{\left(\omega_c+
\mbox{sgn}\left[\frac{q^2}{2m}-\mu\right]\left(\frac{q^2}{2m}-i \omega-\mu
\right)\right)} \;. \nonumber 
\end{eqnarray}
Here, $ \omega_c $ is given by $ B/m $. $ \mbox{sgn}[\cdot] $ is the sign of the
argument. 
We see from this self energy formula that the prefactor of the
$ \log(c) $ term gets
a non trivial frequency dependence. We should mention that the asymptotic
behaviour (\ref{9560}) of the RPA self energy is also correct
in the case of taking into account the Coulomb interaction between the
electrons.

As in the case of the Hartree-Fock self energy, we will calculate
in the following the
Fourier time transformation of $ \Sigma^{\tiny \mbox{RPA}}$. 
For doing this, we have to solve the equation   
$ i\omega-\frac{q^2}{2m}+\mu-\Sigma^{\tiny \mbox{RPA}}=0 $, 
Trepresenting a quadratic equation in the frequencies.
The two frequency solutions correspond to two additive terms of  the
Fourier transform of $\Sigma^{\tiny \mbox{RPA}} $.  
One of the
solutions of this quadratic equation is given by  
$ i \omega=(2\,n_F(q)-1)\, \tilde{\phi}^2 \, \mu
\log(c/k_F)$. 
This solution corresponds to a term in the Fourier transformed RPA Green's
function  which has the correct asymptotics of the exact Green's function
for $ c \to 0$. The other solution is finite for $ c \to 0 $. Thus the
corresponding additive term in the RPA Green's function is finite
for $ c \to 0 $ (modulo logarithmis singularities). Summarizing, the RPA Green's function has not the  
asymptotic behaviour of the exact Chern-Simons Green's function
for $ A \to \infty $.
\\

So far, we did not take into account the self consistence of the
approximation of the Green's function.
Usually  one has to use a self consistent approximation of the 
Green's function in a perturbation theory (e.g. \cite{ne1}). Thus,
we will calculate in the following the self consistent Hartree-Fock Green's function in the leading
$c $ order. The self consistent Hartree-Fock self energy in the leading $ c $
order is a solution of the
following equation 
\begin{eqnarray}
& & \Sigma^{\tiny \mbox{HF}}_{sc}(q) =   
\frac{\tilde{\phi}^2}{4}\frac{k_F^2}{m} \Bigg[\log\left(\frac{c}{k_F}\right) 
n_F(q,\Sigma^{\tiny \mbox{HF}}_{sc}(q)) \nonumber \\
& & -\log\left(\frac{c}{k_F}\right)
 \left(1-n_F(q,\Sigma^{HF}_{sc}(q)) \right)\Bigg]
+\Sigma^{f}_{sc}(q) \;.\eqlabel{2140}
\end{eqnarray}
Here $n_F(q,\Sigma^{\tiny \mbox{HF}}_{sc}) $ is the fermi factor
$ n_F(q,\Sigma^{\tiny \mbox{HF}}_{sc})=1/(\exp[\beta(q^2/(2m)+
\Sigma^{\tiny \mbox{HF}}_{sc}(q)-\mu)]+1)$.
$ \Sigma^{f}_{sc}(q) $ is for $ c \to 0 $ the finite part of the
Hartree-Fock self energy calculated with the self consistent Hartree-Fock
Green's function
$ G^{\tiny \mbox{HF}}_{sc}(\vec{q},\omega)=
-1/(i\omega-q^2/(2m)-\Sigma^{\tiny \mbox{HF}}_{sc}(q)+\mu) $.  
In the leading $ c $ order this self consistent equation is solved by
($T=0$)
\begin{equation}\eqlabel{2150}
\Sigma^{\tiny \mbox{HF}}_{sc}(q,k_F^*) = 
\frac{\tilde{\phi}^2}{4}\frac{(k^*_F)^2}{m}\log\left(\frac{c}{k^*_F}\right)
\left(
2\, \Theta(q-k_F) -1 \right)  
\end{equation}
provided that the fermi momentum $k^*_F $ solves
the following equation
\begin{equation}
\frac{(k_F^*)^2}{2m}-\mu+\Sigma^f(k^*_F,k^*_F)=0 \;. \eqlabel{2160}
\end{equation}
In order to fix $ k^*_F $, we do not have to  calculate  
the finite part $ \Sigma^f(q,k^*_F) $ of the Hartree-Fock self energy. By
using the equation $ 
\lim\limits_{\beta \to \infty}\beta \int d^2k\;n_F(k)\, (1-n_F(k))F(k)
= \int d^2 k\; \delta(|k|-k_F) \, F(k)  $ we get that
 $ \Sigma^f(k^*_F,k^*_F) $ is given by 
\begin{equation} 
\Sigma^f(k_F^*,k_F^*)=\frac{(2\pi)}{m} 
\frac{\partial}{\partial \mu} U^{\tiny \mbox{HF}}((k_F^*)^2/(2m),B)-
\left(\frac{(k_F^*)^2}{2m}\right).
\eqlabel{2330}
\end{equation}
$ U^{\tiny \mbox{HF}}(\mu,B) $ is the Hartree-Fock energy of the Chern-Simons
Hamiltonian (\ref{1050}) (containing the  kinetic energy).
We employed in  (\ref{2330}) the mathematic notation for the 
ordering of the derivation and insertion of the arguments of the functions. 
This means for  equation (\ref{2330}) that we have to partially 
derivate at first the
function $ U^{\tiny \mbox{HF}}(\mu,B)$  depending on the variables $ (\mu,B)$. 
Afterwards 
we have to insert the expressions given in the function brackets.
$ B $ is the external magnetic field.
The Hartree-Fock energy of the $ \nu=1/\tilde{\phi} $ system
is given by \cite{sit1,di3}
\begin{equation} \eqlabel{2315}
U^{\tiny \mbox{HF}}(\mu,\tilde{\phi} m \mu)=\frac{m}{4 \pi}
\mu^2+\frac{3 m }{16 \pi} \tilde{\phi}^2 \mu^2  \,.
\end{equation}
Since we calculate $ U^{\tiny \mbox{HF}}(\mu,B) $ in (\ref{2315})
for $ B=\tilde{\phi} m \mu $ it is not correct to insert  
(\ref{2315}) in (\ref{2330}). This was shown in \cite{di2}
generally in the case of the
determination of $ \mu $ through the equation
$-\partial \Omega/(\partial \mu) =N/A $.
Here $ \Omega $ is the grand canonical potential of the Chern-Simons system.  
We obtained  in the paper \cite{di2} that one
gets a correction to this equation if $ \Omega $ is calculated under the
constraint $ B=2 \pi \tilde{\phi} N/A $.
With the help of the derivations in this paper it is easy to see that
the following equation  
results in the correct $ k_F^* $ 
\begin{equation} 
\Sigma^f(k_F^*,k_F^*)\nts{1} = \nts{1} \frac{(2\pi)}{m} 
\frac{\partial}{\partial \mu}
U_{\nu=1/\tilde{\phi}}^{\tiny \mbox{HF}}\nts{2}\left(\frac{(k_F^*)^2}{2m}\right)+\Sigma_c(k_F^*)-
\left(\nts{1} \frac{(k_F^*)^2}{2m} \nts{1} \right)   
\eqlabel{2345}
\end{equation}
Here $ U_{\nu=1/\tilde{\phi}}^{\tiny \mbox{HF}}(\mu) $ is given by
$ U_{\nu=1/\tilde{\phi}}^{\mbox{\tiny \mbox{HF}}}(\mu):=
U^{\tiny \mbox{HF}}(\mu,\tilde{\phi} m \mu) $.
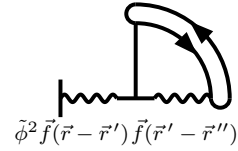
\begin{figure}[t]
\begin{center}  
\begin{pspicture}(6,2.3)\psset{unit=0.5cm}
\psset{linewidth=1.5pt,arrowinset=0}
\psarcn{->}(5,2){2.0}{90}{35}
\psarcn{-}(5,2){2.0}{45}{0}
\WavyLine[n=25,beta=180](3,2)(4.5,2)
\WavyLine[n=25,beta=180](5.5,2)(7,2)
\psline[ArrowInside=-](4.5,2)(5.5,2)
\psline[ArrowInside=-](5,2)(5,4) 
\psline[ArrowInside=-](3,1.5)(3,2.5)
\psarcn{-}(5,4.25){0.25}{-90}{90}
\psarcn{-}(7.25,2){0.25}{0}{-179}
\psarcn{-}(5,2){2.5}{90}{45}
\psarcn{<-}(5,2){2.5}{55}{0}
\rput(6.3,1.1) {\footnotesize $ \vec{f}(\vec{r}\,'-\vec{r}\,'')$}
\rput(3.3,1.1) {\footnotesize $\tilde{\phi}^2 \vec{f}(\vec{r}-\vec{r}\,')$}
\end{pspicture}
\caption{The self energy graph $ \Sigma_c $. \label{bildselb}}
\end{center}
\end{figure}
The self energy diagram $ \Sigma_c $ is shown in figure \ref{bildselb}.
By closing the open ends of the self energy diagram by an
interaction free Green's function
we see that the resulting $ \Omega $ diagram is canceled under the condition
$B =\tilde{\phi}m\mu  $ with a diagram where this Green's function is
replaced by a coupling to the external magnetic field $B$. 
Furthermore, we see from the figure that the momentum transfer to
the external coupling is zero for an infinite system. Thus, we get difficulties
in calculating this diagram (because  $ \lim_{q \to 0} \vec{f}(q)=\infty $).  
We showed in \cite{di2} that the correct way to calculate this diagram is
to calculate the integrals of the diagram for a finite system taking 
at last the limit $ A \to \infty $.  
By doing this we get  
\begin{eqnarray}
\lefteqn{\Sigma_c(q) =- \frac{1}{m}\frac{(2\pi\tilde{\phi})^2}
{(2\pi)^4}}\eqlabel{2360} \\
& & 
\lim\limits_{q \to 0} 
\int d^2k_1 d^2k_2 n_F(|\vec{k}_1-\vec{k}_2|)
n_F(|\vec{k}_2-\vec{q}|) \frac{\vec{k}_1}{|\vec{k}_1|^2}
\frac{\vec{q}}{|\vec{q}|^2} . \nonumber 
\end{eqnarray}
To calculate the double integral we expand $ n_F(|\vec{k}_2-\vec{q}|) $
for small $q $ and $T>0 $. By carrying out the integration
we get 
\begin{equation} \eqlabel{2370}
\Sigma_c(q)=-\tilde{\phi}^2 \mu  \,.
\end{equation}
By inserting  (\ref{2315}) and (\ref{2370}) in equation (\ref{2345}) 
we get with the help of (\ref{2160}) 
\begin{equation}\eqlabel{2380}
\frac{(k^*_F)^2}{2m}=\frac{\mu}{(1+\frac{\tilde{\phi}^2}{4})}  \;.
\end{equation}
In the following, we will make a similar calculation
neglecting $ \Sigma_c(q) $ in
(\ref{2345}). We denote $ \tilde{k}_F^* $
by the fermi momentum which is
calculated through the equations (\ref{2160}) and (\ref{2345})
with $ \Sigma_c(k)=0 $. This fermi momentum is given by 
\begin{equation}  \eqlabel{2340}
\frac{(\tilde{k}_F^*)^2}{2m}=\frac{\mu}{(1+\frac{3}{4}\tilde{\phi}^2)} \;.
\end{equation}

Now we calculate the density of the system by using the self-consistent
Hartree-Fock Green's function. 
With the help of 
$ G^{\tiny \mbox{HF}}_{sc}(\vec{q},\omega,k_F^*)=-1/(i\omega-q^2/(2m)-\Sigma^{\tiny \mbox{HF}}_{sc}
(q,k_F^*)+\mu) $
we get for the electron density
$ N/A=\sum_{\omega} \sum_{\vec{q}} G^{\tiny \mbox{HF}}_{sc}(\vec{q},\omega,k_F^*)
e^{i \omega \eta} $
($ \eta $ is an infinite small positive parameter) with $ \mu= (2 \pi) m N/A $
for the $ \nu=1/2 $ system we get $1/2 $ of the electron density.
In contrast to this  the trace over the exact Green's function
is given by the electron density. 
In the paper \cite{di2}, we showed  that
$  -\partial \Omega '(\mu,\infty )/(\partial \mu)=0 $ . $ \Omega'(\mu,\beta) $ 
is the grand canonical potential of the $ \nu=1/\tilde{\phi} $ system
which is calculated under the constraint $ B= (2\pi \tilde{\phi} )N/A $.
From this equation it is clear that
$ \sum_{\omega} \sum_{\vec{q}} G^{\tiny \mbox{HF}}_{sc}(\vec{q},\omega,\tilde{k}_F^*)
e^{i \omega \eta} $ should be zero to be a good approximation.
For the $ \nu=1/2 $ system, we get for this
expression with
$ \mu= (2 \pi) m N/A $,  $ 1/4 $  of the
electron density. Thus, we get the correct  relation between the densities
calculated with the help of the Hartree-Fock Green's function 
  $ G^{\tiny \mbox{HF}}_{sc}(\vec{q},\omega,k_F^*) $ and
 $ G^{\tiny \mbox{HF}}_{sc}(\vec{q},\omega,\tilde{k}_F^*)$. 
When formulating a perturbation theory for the grand canonical
potential which uses the Hartree-Fock Green's function as the mean field
Green's function we have to use 
$ G^{\tiny \mbox{HF}}_{sc}(\vec{q},\omega,\tilde{k}_F^*) $
because (as discussed above) insertions of the self energy $ \Sigma_c $
in $ \Omega $ diagrams are cancelled by $ B $ coupling diagrams. The 
formulation of this theory will be published in a later paper.

Until now we did not take into account the Coulomb interaction.
By this interaction we obtain in the Hartree-Fock approximation an additional
Fock diagram including the Coulomb vertex. This Coulomb diagram is finite.
Thus we get also in this case an asymptotic Green's function of the
form (\ref{2180}). This is also in agreement with the results of 
subsection II E. 
At last, we have to mention that we obtain a difference for
$\Sigma_c $   from the result of Sitko and Jacak \cite{sit2}
(they calculated  $\Sigma_c=0 $). 
The reason is that they did not take
the limit $ A \to \infty $ at the end of the calculation of $\Sigma_c $.

\section{Conclusion}
In this paper, we calculated the asymptotic form of the
$ \nu=1/\tilde{\phi} $ Chern-Simons Green's function for an
infinite area $ A $ non-perturbationally. This was done concretely
for the Coulomb free $ \nu=1/\tilde{\phi} $ Chern-Simons theory. We obtain
that the asymptotics of the Green's function behaves as
$ G(\vec{r},\tau) \sim e^{-f(r,\tau)\log(A)} $.
Due to the sign of $ \tau $ the function $ f(r,\tau) $ can be written as two
different power expansions in $ \tau $. 
We calculated $ f(r,\tau) $
to the third order in $ \tau $. Due to this calculation
we get that $ f(r,\tau) $
results in a positive function. It would be
interesting to see if
this is also true considering higher powers of $ \tau $. 
Next, we discussed the asymptotic behaviour of the Green's function
for the Chern-Simons
theory taking into consideration the Coulomb interaction. We obtain
(for temperature $ T>0 $)  the same asymptotics as for the Green's function
of the Chern-Simons
theory without Coulomb interaction (in this case $ f(r,\tau) $
depends also on the  Coulomb coupling constant $ e^2 $).

In section III, we examined the Green's function of the Chern-Simons theory in
the Hartree-Fock approximation as well as in the RPA without Coulomb interaction. We obtained that the
asymptotics of the Hartree-Fock approximation of the Green's function
behaves for $ \vec{r}=0 $ and
small $ \tau $ similar to the asymptotics of the exact one. Especially the
prefactor of the $ \log(A) $ term of the Hartree-Fock Green's
function is the same as of the exact Green's function.
Next, we calculated the asymptotic behaviour
of the RPA
Green's function. We showed that this Green's function is finite for
$ A \to \infty $. This is not in correspondence with the exact Green's
function. 
On the way to formulate a perturbation theory around the Hartree-Fock
mean field, we examined the self consistent
Hartree-Fock Green's function.
We solved the self consistence equation in the leading $ 1/A $ order.
We obtain 
a self consistent Hartree-Fock Green's function which behaves similar
to the Hartree-Fock Green's function with the difference that it has a
different fermi momentum. We obtained that the density of the electrons
calculated with the help
of the self consistent Hartree-Fock Green's function is $ 1/2 $ of the
exact density (for the $ \nu=1/2 $ system).
Furthermore, we calculated the self consistent Green's function without one of
the Hartree-Fock self energy diagrams which is zero when inserted in
$ \Omega $ diagrams. The electron density which is calculated with the help
of this Green's function is $ 1/4 $ of the exact electron density (for the $
\nu=1/2 $ system). It was shown by us in \cite{di2} that 
the electron density calculated with this truncated Green's function should be
zero. At last we obtained that the asymptotics of the Green's function
in the Hartree-Fock approxmation by taking into account
the Coulomb interaction has the same form as the Green's function
without Coulomb interaction. This is in accordance with the exact results
of section II. 

By taking the asymptotics of the Green's function
seriously with respect to the principles of perturbational many-body theory 
we now have two options to go further. First, we
may establish a theory  which integrates the $\log(A) $ singularity by using
the Hartree-Fock Green's function
as the mean field Green's function. This theory was formulated by us
in \cite{di3}. It is our purpose to publish the results in a subsequent paper.
Second, we may establish other formulations of the Chern-Simons theory
(i.e. \cite{sh1,pas1}) in the hope to get a well behaved Green's function.
One may think that this could be reached by the theory of Shankar and Murthy
\cite{sh1} speculated earlier. 
This assumption is in contrast to \cite{st2} where it was shown   
in RPA that one gets also a singular self energy in this theory
(for temperature $ T=0 $). 

\bigskip
We would like to thank K. Luig and W. Weller for many helpful 
discussions during the course of this work. 
Further we have to acknowledge the financial support by the Deutsche
Forschungsgemeinschaft, Graduiertenkolleg "Quantenfeldtheorie".

\begin{appendix}
\section{ The commutators for the calculation of the Chern-Simons Green's
  function} 
With the help of the equation (\ref{1620}) we get from (\ref{1530})
to order $ \tau^3 $
cumulant expectation values of the operator 
$ H_{N+1}+\overline{H}_{N+1} $ of the form $ 
\sum_{u_p \in u_{0,\vec{k}}} 1/p +O(1/p^2) $ and
$ \sum_{u_p,u_{p+1}\in u_{0,k}} 1/p +O(1/p^2) $. 
The terms of the form $ \sum_{u_p \in u_{0,\vec{k}}} 1/p+O(1/p^2) $
are given by
(we write down only
these terms of the expectation value which scales
as $ \sum 1/p $).  
 \begin{eqnarray}
\left<u_{0,\vec{k}}\left|\,\left[H_{4,1},\left[H_{0,1},H_{4,1}\right]\right]\,
\right|u_{0,\vec{k}}\right>_{c} & \nts{2}=\nts{1} & 
\tilde{\phi}^2 \frac{1}{2}\sum_{u_p \in
  u_{0,\vec{k}}}\frac{1}{p}  , \nonumber\\
\left<u_{0,\vec{k}}\left|\,\left[H_{0,1},H_{4,1}\right](H_{4,1}+H_3)\,
\right|u_{0,\vec{k}}\right>_{c} & \nts{2}=\nts{1} & 
-\tilde{\phi}^2 \frac{1}{4}\sum_{u_p \in u_{0,\vec{k}}}\frac{1}{p},\nonumber\\
\left<u_{0,\vec{k}}\left|\,(H_{4,1}+H_3) \left[H_{0,1},H_{4,1}\right]
\right|u_{0,\vec{k}}\right>_{c} & \nts{2}=\nts{1} & 
\tilde{\phi}^2 \frac{1}{4}\sum_{u_p \in
  u_{0,\vec{k}}}\frac{1}{p},\nonumber\\
\left<u_{0,\vec{k}}\left|\,\left[H_{4,2},\left[H_{0,1},H_{4,1}\right]\right]
\,\right|u_{0,\vec{k}}\right>_{c} & \nts{2}=\nts{1} & 
\tilde{\phi}^2 \frac{3}{8}\sum_{u_p \in u_{0,\vec{k}}}\frac{1}{p},\nonumber\\
\left<u_{0,\vec{k}}\left|\,H_{4,2} \left[H_{0,1},H_{4,1}\right]
\,\right|u_{0,\vec{k}}\right>_{c} & \nts{2}=\nts{1} & 
\tilde{\phi}^2 \frac{3}{8}\sum_{u_p \in u_{0,\vec{k}}}\frac{1}{p} ,\nonumber \\
\left<u_{0,\vec{k}}\left|\,\left[H_{4,1},\left[H_{0,2},H_{4,1}\right]\right]\,
\right|u_{0,\vec{k}}\right>_{c} & \nts{2}=\nts{1} & 
\tilde{\phi}^2 \frac{1}{8}\sum_{u_p \in
  u_{0,\vec{k}}}\frac{1}{p} ,  \nonumber\\
\left<u_{0,\vec{k}}\left|\,\left[H_{0,2},H_{4,1}\right](H_{4,1}+H_3)
\,\right|u_{0,\vec{k}}\right>_{c} 
& \nts{2}=\nts{1} & 
-\tilde{\phi}^2 \frac{1}{8}\sum_{u_p \in
  u_{0,\vec{k}}}\frac{1}{p},\nonumber\\
\left<u_{0,\vec{k}}\left|\,\left[H_{4,1},\left[H_{0,2},H_{4,2}\right]\right]\,
\right|u_{0,\vec{k}}\right>_{c} & \nts{2}=\nts{1} & 
\tilde{\phi}^2 \frac{1}{8}\sum_{u_p \in u_{0,\vec{k}}}\frac{1}{p},\nonumber\\
\left<u_{0,\vec{k}}\left|\,\left[H_{0,2},H_{4,2}\right](H_{4,1}+H_3)\,
\right|u_{0,\vec{k}}\right>_{c} & \nts{2}=\nts{1} & 
-\tilde{\phi}^2 \frac{1}{8}\sum_{u_p \in u_{0,\vec{k}}}\frac{1}{p}
,\nonumber\\
\left<u_{0,\vec{k}}\left|\,\left[H_{4,2},\left[H_{0,2},H_{4,1}\right]\right]\,
\right|u_{0,\vec{k}}\right>_{c} & \nts{2}=\nts{1} & 
\tilde{\phi}^2 \frac{1}{8}\sum_{u_p \in u_{0,\vec{k}}}\frac{1}{p},\nonumber\\
\left<u_{0,\vec{k}}\left|\,H_{4,2}\left[H_{0,2},H_{4,1}\right]\,
\right|u_{0,\vec{k}}\right>_{c} & \nts{2}=\nts{1} & 
\tilde{\phi}^2 \frac{1}{8}\sum_{u_p \in u_{0,\vec{k}}}\frac{1}{p}
,\nonumber\\
\left<u_{0,\vec{k}}\left|\,\left[H_{4,2},\left[H_{0,2},H_{4,2}\right]\right]
\,\right|u_{0,\vec{k}}\right>_{c} & \nts{2}=\nts{1} & 
\tilde{\phi}^2 \frac{1}{8}\sum_{u_p \in u_{0,\vec{k}}}\frac{1}{p},\nonumber\\
\left<u_{0,\vec{k}}\left|\,H_{4,2} \left[H_{0,2},H_{4,2}\right]
\,\right|u_{0,\vec{k}}\right>_{c} & \nts{2}=\nts{1} & 
\tilde{\phi}^2 \frac{1}{8}\sum_{u_p \in u_{0,\vec{k}}}\frac{1}{p}.\nonumber\\
\eqlabel{60000}     
\end{eqnarray}
We get the remaining terms  of the form $  \sum_{u_p,u_{p+1}\in u_{0,k}}
1/p+O(1/p^2) $ of the cumulant expectation value of the operator
$ H_{N+1}+\overline{H}_{N+1} $ by
dropping the first three terms in the above list and substituting
$ \sum_{u_p \in u_{0,\vec{k}}} $ by  $ -\sum_{u_p,u_{p+1} \in u_{0,\vec{k}}}$.  
\end{appendix}

\end{multicols}


\begin{thebibliography}{10}

\bibitem {ts1}
D.C. Tsui, H.L. St\"ormer, and A.C. Gossard, Phys. Rev. Lett. {\bf 48}, 1559
 (1982); D.C. Tsui, H.L. St\"ormer, and A.C. Gossard, Phys. Rev. B {\bf
 25}, 1405 (1982).

\bibitem{ja1}
J.K. Jain, Phys. Rev. Lett. {\bf 63}, 199 (1989).

\bibitem{hlr}
B.I. Halperin, P.A. Lee, and N. Read, Phys. Rev. B {\bf
 47}, 7312 (1993).

\bibitem{ka1}
 V. Kalmeyer, and S.C. Zhang, Phys. Rev. B {\bf 46}, 9889 (1992).
 
\bibitem{kan1}
W. Kang et al., Phys. Rev. Lett. {\bf 71}, 3850 (1993).

\bibitem{sm1}
J.H. Smet et al., Phys. Rev. Lett. {\bf 77}, 2272 (1996).

\bibitem{wil2}
R.L. Willet, Adv. Phys. {\bf 46}, 447 (1997).

\bibitem{st1}
A. Stern, and B.I. Halperin, Phys. Rev. B {\bf 52}, 5890 (1995).

\bibitem {sh1}
R. Shankar, and G. Murthy, Phys. Rev. Lett. {\bf 79}, 4437 (1997).

\bibitem {st2}
A. Stern, B.I. Halperin, F.v. Oppen,
and S.H. Simon, Phys. Rev. B {\bf 59}, 12547 (1999).

\bibitem{pas1}
V. Pasquier, and F. D. M. Haldane, Nucl. Phys. B {\bf 516}, 719
(1998); N. Read, Phys. Rev. B {\bf 58}, 16262 (1998); D.-H. Lee, Phys. Rev. Lett {\bf 80}, 4547 (1998)

\bibitem{sit2}
P. Sitko, and L. Jacak, Mod. Phys. Lett. B. {\bf 9}, 889 (1995).

\bibitem{sim1}
S.H. Simon in {\em Composite Fermions}, Ed. O. Heinonen,
World Scientific, Singapore (1998).

\bibitem{ha2}
B. I. Halperin in {\em Perspectives in QHE}, Eds. Das Sarma and Pinczuk,
J. Wiley, NY (1997).

\bibitem{zh2}
S.C. Zhang, Int. J. Mod. Phys. B {\bf 6}, 25 (1992).

\bibitem {mo1}
R. Morf, and N. d'Ambrumenil, Phys. Rev. Lett. {\bf 74}, 5116 (1995).

\bibitem{di1}
J. Dietel, Eur. Phys J. B {\bf 19}, 195 (2001).

\bibitem{ne1}
J.W. Negele, and H. Orland, {\em Quantum Many-Particle Systems},
Addison-Wesley New York (1994).

\bibitem{sit1}
P. Sitko, Phys. Lett. A {\bf 188}, 179 (1994).

\bibitem{di3}
J. Dietel, Ph. D. thesis, University of Leipzig (2000) (unpublished).

\bibitem{di2}
J. Dietel, Eur. Phys. J. B {\bf 22}, 43 (2001).

\end{thebibliography}
\end{document}